\documentclass[12pt]{article}
\usepackage{amsmath}
\usepackage{amssymb}
\usepackage{epsfig}
\usepackage{lscape}

\begin{document}

\begin{center}
{\Large \bf Value of \boldmath{$\alpha_{\rm s}$} from 
deep-inelastic-scattering data}
\vspace{1cm}

{\bf S.~I.~Alekhin}
\vspace{0.1in}

{\baselineskip=14pt Institute for High Energy Physics, 142281 Protvino, Russia}
\begin{abstract}
We report the value of $\alpha_{\rm s}$ obtained from QCD analysis 
of existing data on deep-inelastic scattering
of charged leptons off proton and deuterium and 
estimate its theoretical uncertainties with
particular attention paid to impact of the high-twist contribution 
to the deep-inelastic-scattering structure functions.
Taking into account the major uncertainties the value 
$\alpha^{\rm NNLO}_{\rm s}(M_{\rm Z})=0.1143
\pm 0.0014({\rm exp.})\pm 0.0013({\rm theor.})$ is obtained. 
An extrapolation of the LO--NLO--NNLO results to the higher orders
makes it possible to 
estimate $\alpha^{\rm N^3LO}_{\rm s}(M_{\rm Z})\sim 0.113$.
\end{abstract}
\end{center}
{\bf PACS numbers:} 06.20.Jr,12.38.Bx\\
{\bf Keywords:} deep inelastic scattering, strong coupling constant

\newpage

\section{Introduction} 
The value of strong coupling constant $\alpha_{\rm s}$ can be extracted 
from the existing data on deep-inelastic-scattering (DIS)
with experimental uncertainty at the one-percent level. 
With such experimental accuracy achieved the theoretical uncertainties
become dominant. One of the major theoretical uncertainty
in $\alpha_{\rm s}$ derived  
from the next-to-leading-order (NLO) QCD analysis of DIS data comes from 
the higher-order (HO) corrections. 
Therefore, consideration of the next-to-next-to-leading (NNLO)
corrections is of great importance
for suppression of the errors in $\alpha_{\rm s}$. 
Indeed, account of the NNLO terms in the 
QCD evolution equations for the DIS structure functions, which 
have been calculated recently~\cite{Retey:2000nq,vanNeerven:2000wp},
allows to reduce the HO uncertainty 
in $\alpha_{\rm s}$ by a factor of $2\div 3$,
as it was estimated in Ref.~\cite{vanNeerven:1999ca}. Another important
source of the uncertainty in $\alpha_{\rm s}$ is the high-twist (HT)
contribution to the structure functions. 
This contribution inevitably appear in the 
operator product expansion~\cite{Wilson:zs}, but unfortunately 
cannot be reliably estimated hitherto. In the limited range 
of the momentum transfer $Q$ the HT terms, which fall with $Q$ 
by the power law, can simulate the logarithmic-type behavior of the 
leading twist (LT) term. 
For this reason the magnitude of HT terms is correlated 
with the value of $\alpha_{\rm s}$, which defines the slope 
of the LT term, and the uncertainty in HT terms propagates into 
the uncertainty of $\alpha_{\rm s}$. 

\begin{figure}[ht]
\centerline{\epsfig{file=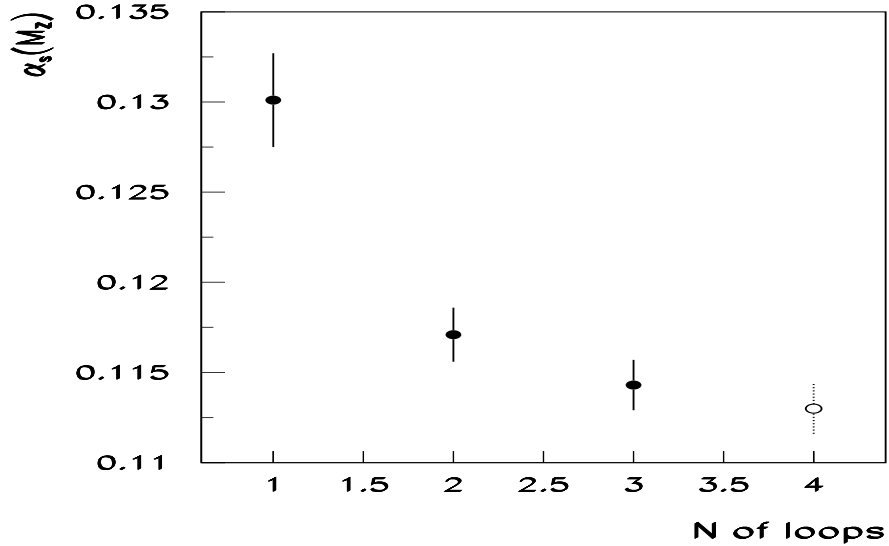,width=14cm,height=8cm}}
\caption{The central 
values of $\alpha_{\rm s}(M_{\rm Z})$ and the experimental errors
obtained in different orders of pQCD (full symbols). The open symbol shows
the result of extrapolation to the N$^3$LO.}
\label{fig:als}
\end{figure}

We study impact of these 
sources of uncertainties on the value of $\alpha_{\rm s}$ obtained
from a NNLO QCD analysis of existing data on the charged-leptons
DIS off the proton and deuterium targets~\cite{Whitlow:1992uw}.
The data outside the kinematical region of
$Q^2=2.5\div 300~{\rm GeV}^2$ and $x=10^{-4}\div 0.75$ are left out
to suppress potentially dangerous theoretical uncertainties 
keeping sufficient sensitivity of 
data to the value of $\alpha_{\rm s}$. The data are analyzed   
using a numerical integration of the QCD evolution equations in 
the $x$-space (details are described elsewhere~\cite{Alekhin:2001ch}).
The value of $\alpha_s$ obtained in the recent update of this fit
is~\cite{Alekhin:2002fv}
\begin{equation}
\alpha^{\rm NNLO}_{\rm s}(M_{\rm Z})=0.1143\pm 0.0014({\rm exp.}),
\label{eqn:asa}
\end{equation}
to be compared with
\begin{equation}
\alpha^{\rm NNLO}_{\rm s}(M_{\rm Z})=0.1166\pm 0.0009({\rm exp.})
\label{eqn:asy}
\end{equation}
obtained by Santiago and Yndurain from the analysis of the proton subset of 
data~\cite{Santiago:2001mh}.
The achieved experimental error in $\alpha_s$ is quite small and
therefore, complimentary analysis of theoretical 
uncertainties is necessary in order
to clarify the confident range of $\alpha_{\rm s}$ variation.

\section{Uncertainty due to the higher-order corrections} 

Generally, estimate of the HO uncertainty is impossible 
before the HO corrections are known. However, the approach 
based on variation of the QCD renormalization scale (RS) is 
commonly used for estimation of the HO uncertainties.
This approach is very approximate since it does not account 
for a possible $x$-dependence of the corresponding scale-factor
and since the variation range of the scale-factor
is optional. Nevertheless, following this
approach we estimate the HO uncertainty as the change of $\alpha_{\rm s}$ 
caused by variation of the QCD RS from $Q$ to $2Q$.
The estimates obtained are given in Table~\ref{tab:als}.
Note the reduction of $\alpha_{\rm s}$ with the pQCD order
approximately coincide with the corresponding 
HO uncertainties\footnote{The decrease in
$\alpha_{\rm s}$ because of going 
from the NLO to the NNLO disagrees 
with the results of NNLO analysis by Santiago and Yndurain, but 
agrees with the estimates of Ref.\cite{vanNeerven:1999ca}
and results of the NNLO fit of Ref.\cite{Martin:2002dr}.}. 
Extrapolating this regularity to the N$^3$LO we have an estimate
$\alpha_{\rm s}^{\rm N^3LO}(M_{\rm Z})\sim 0.113$.

\begin{table}[ht]
\begin{center}
\caption{The values of $\alpha_{\rm s}(M_{\rm Z})$ obtained in different 
orders of pQCD analysis and their RS errors.} 
\vspace{0.4cm}
\begin{tabular}{|c|c|}
\hline
LO&   $0.1301\pm0.0026 ({\rm exp.})\pm0.0149 ({\rm RS})$ \\ \hline
NLO & $0.1171\pm0.0015 ({\rm exp.})\pm0.0033 ({\rm RS})$ \\ \hline
NNLO & $0.1143\pm0.0014 ({\rm exp.})\pm0.0009 ({\rm RS})$ \\ \hline
\end{tabular}
\end{center}
\label{tab:als}
\end{table}

\section{Uncertainty due to the high-twist contribution} 

The HT terms can be taken into account using different approaches.
Firstly, one can parameterize the HT terms by a flexible 
model independent function. In this approach possible uncertainties 
in variation of the HT terms allowed by data merge into the 
total experimental error in $\alpha_{\rm s}$.
Secondly, one can cut data with low 
momentum transfer $Q$ and/or hadronic mass $W$. 
Evidently in the second approach the HT error vanishes, however 
the statistical error can rise 
because the QCD evolution introduces the largest effect 
at small $Q$ and therefore this region 
is most sensitive to the value of $\alpha_{\rm s}$. 
Thirdly, one can take into account the HT terms using 
available theoretical models. In this approach the HT uncertainty 
is subject of belief in reliability of the model.

\begin{figure}[ht]
\centerline{\epsfig{file=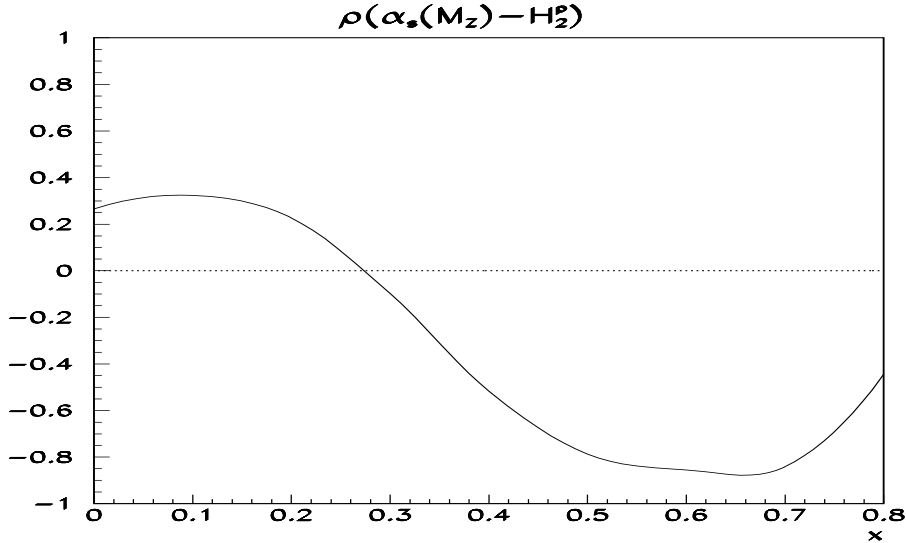,width=14cm,height=8cm}}
\caption{Shown is the correlation coefficient for the 
fitted values of $\alpha_{\rm s}(M_{\rm Z})$ and $H_2^p$
at different $x$.}
\label{fig:cor}
\end{figure}

\begin{figure}[ht]
\centerline{\epsfig{file=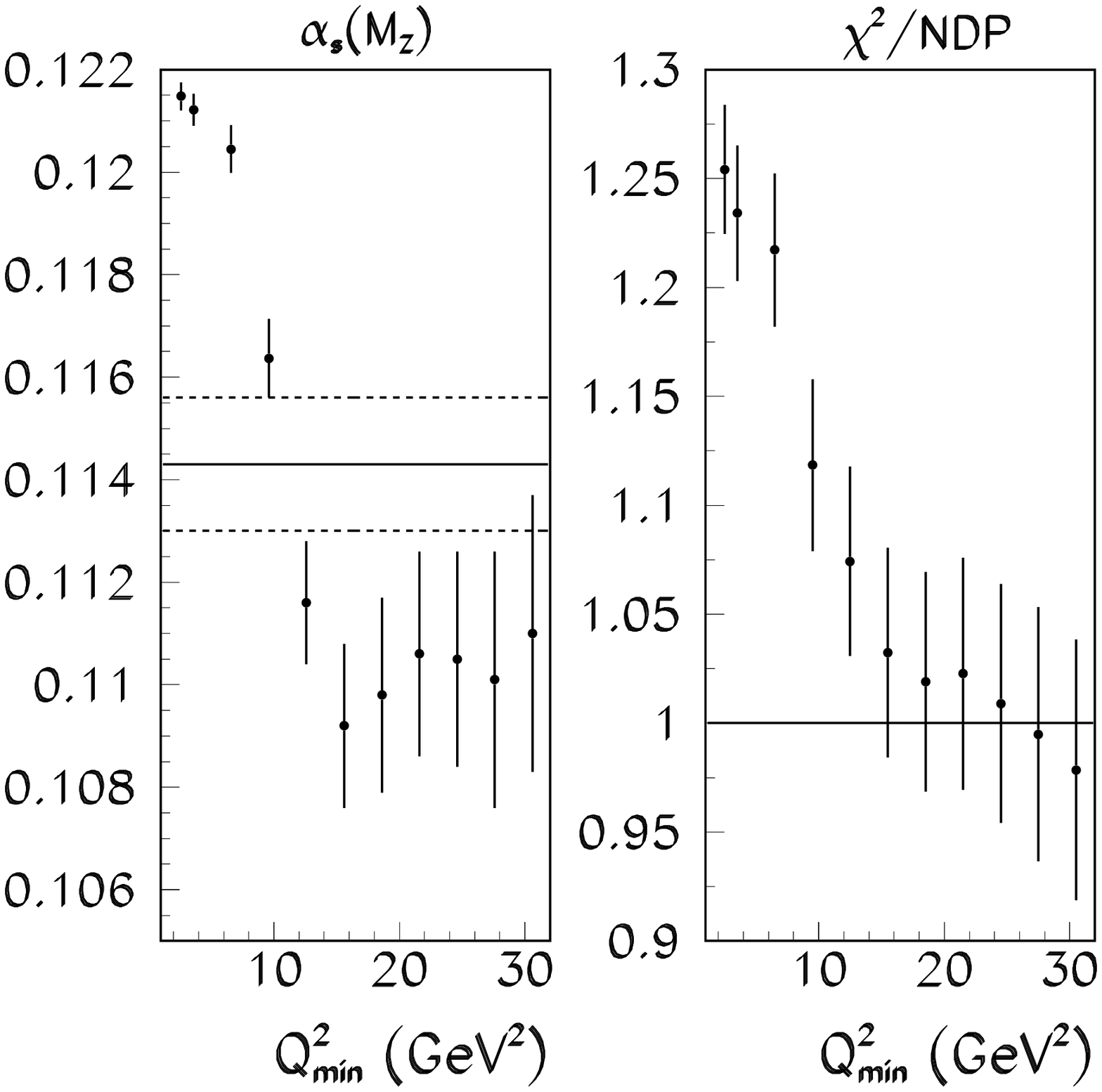,width=13cm,height=8cm}}
\caption{The values of $\alpha_{\rm s}(M_{\rm Z})$ obtained from the 
fit with $H_{\rm 2,L}=0$ and different cuts $Q_{\rm min}$ (left panel)
and the corresponding 
values of $\chi^2/NDP$ (right panel). The horizontal lines
in the left panel correspond to the central value and the error band
of $\alpha_{\rm s}(M_{\rm Z})$ in the analysis with 
the HT parameterized in the piece-linear form, the error bars in the 
right panel are $\sqrt{2/NDP}$.}
\label{fig:scanq}
\end{figure}

In the analysis of Ref.\cite{Alekhin:2002fv} we parameterize 
the twist-4 terms in additive form,
$F_{\rm 2,L}=F_{\rm 2,L}^{\rm LT,TMC}+{H_{\rm 2,L}(x)}/{Q^2}$,
where $F_{\rm 2,L}^{\rm LT,TMC}$ are the LT terms corrected  
for the target mass effects~\cite{Georgi:1976ve}. The functions
$H_{\rm 2,L}$ are defined in a piece-linear form with fitted coefficients
that provides the model-independent account of the HT terms.
One does need to include 
the HT terms in the analysis since they improve quality of the fit  
($\chi^2/NDP=2521/2274$ with and $\chi^2/NDP=2851/2274$ without them).
Possible logarithmic $Q$-dependence of $H(x)$ at $x\rightarrow 1$
could decrease the 
fitted $\alpha_{\rm s}$ value~\cite{Gardi:2002xm}, however
data used in our analysis
are not sensitive to such modifications of $H(x)$
because of the cut on $x$. 
The value of $\alpha_{\rm s}$ is strongly anti-correlated with 
$H_2$ at large $x$ (see Fig.\ref{fig:cor}).
For this reason the error in $\alpha_{\rm s}$ is much 
larger than it could be in the absence of the HT terms.
Indeed, in our trial fit with the HT terms set
to 0, the error in $\alpha_{\rm s}(M_{\rm Z})$ is 0.0003.
Comparing this error with the error in $\alpha_{\rm s}$ obtained in our 
analysis with the HT terms accounted for, we estimate the HT error in 
$\alpha_{\rm s}(M_{\rm Z})$ to be 0.0014. This is  the major 
source of the error in $\alpha_{\rm s}^{\rm NNLO}$, more important than 
the estimated uncertainty due to the HO.

The results of scan fits 
with different cuts on $Q$ and with no HT terms included 
are given in Fig.\ref{fig:scanq}.
These results suggest the optimal value  
of $Q_{\rm min}\sim 20~{\rm GeV}^2$, which provide 
balance between stability of the central value of $\alpha_{\rm s}$
and its error. For $Q^2_{\rm min}=21.5~{\rm GeV}^2$  the value of 
$\alpha_{\rm s}(M_{\rm Z})=0.1106\pm 0.0020~({\rm exp.})$  was 
obtained. This value agrees with the results of analysis 
with the HT terms included that supports the both approaches. 
At that the value of $\alpha_s$ derived from the fit with 
the HT terms included is preferable since it is more robust due 
to the wider set of data was used in this case. The cut on $W$ leaves out the 
data at small $Q$ and large $x$ and therefore 
makes fit less sensitive to the HT terms however not so efficiently 
as the cut on $Q$ since 
the HT terms are important for moderate $x$ as well.
With the increase of the minimal value of $W$
the error in $\alpha_{\rm s}$ gets inappreciably large earlier than 
its central value is sustained.

\begin{figure}[ht]
\centerline{\epsfig{file=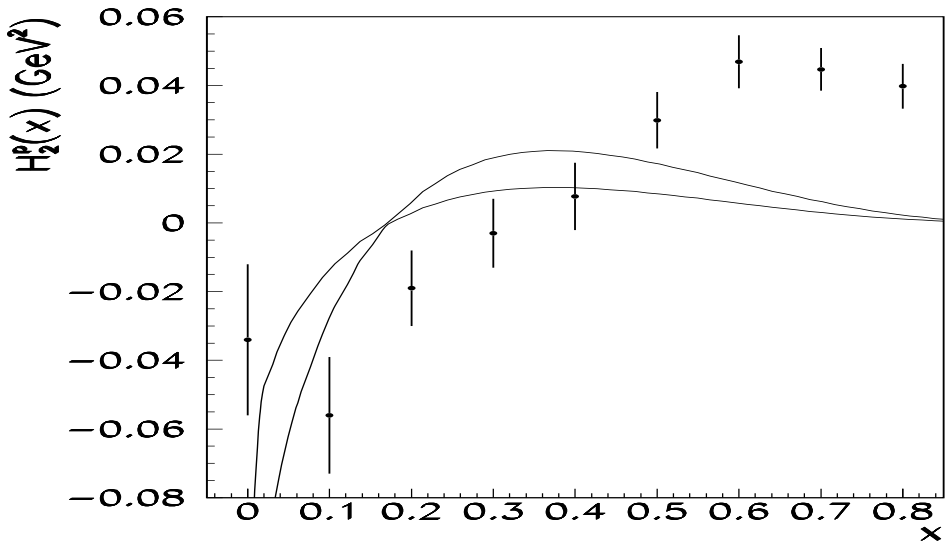,width=14cm,height=8cm}}
\caption{One-standard-deviation band  for the HT term in the proton structure
function $F_2$ obtained in the gradient model fit (curves) compared to the
model independent determination (points).}
\label{fig:gsa}
\end{figure}

In the analysis by Santiago and Yndurain the value of $\alpha_{\rm s}$ was 
obtained from the fit with the HT terms set to 0. The uncertainty in 
$\alpha_{\rm s}$ caused by this constraint was estimated as 
a difference between the value of $\alpha_{\rm s}$ obtained
from the fit with the HT terms set to 0 and that of
the fit with HT terms described by the gradient model. 
In this model the HT contribution to $F_2$ reads
$A_2 [{dF_2^{NS}(x)}/{dx}]/Q^2$,
where $F_2^{NS}$ is the non-singlet structure function and $A_2$ is 
a normalization parameter determined from data.

We find $A_2=-0.018\pm0.006~{\rm GeV}^2$ in the trail 
fit with $H_2(x)$ described by the gradient model\footnote{In 
the trial fit we leave out the data
points with $Q^2<3.5~{\rm GeV}^2$ in order 
to better match the data set used in Ref.\cite{Santiago:2001mh}.}. 
We find that in this fit $\alpha_{\rm s}(M_{\rm Z})$ changes by 0.0006
with respect to that of the fit with HT terms set to 0.
This value agrees with 0.0004 
obtained in the analysis by Santiago and Yndurain.
However this shift in $\alpha_{\rm s}$ cannot be considered 
as a reliable estimate of the HT uncertainty in $\alpha_{\rm s}$, since 
the gradient model describes the HT terms very poorly ($\chi^2/NDP=2543/2067$).
This is not surprising, because the gradient model can only be justified  
at $x\rightarrow 1$  and cannot be applied in
the whole region of $x$ (see Fig.\ref{fig:gsa}).
A correct estimate of the HT uncertainty in $\alpha_{\rm s}$ 
in the analysis by Santiago and Yndurain would be given by a variation  
of $\alpha_{\rm s}(M_{\rm Z})$ due to going
from the fit 
with the HT terms set to 0 to the fit with the HT terms parameterized 
in the model-independent way. In our analysis such variation 
is $-0.0069$, much larger than 
estimate of the HT uncertainty based on the gradient model.

Taking into account this variation, we conclude that in fact 
Eqn.(\ref{eqn:asa}) disagrees with Eqn.(\ref{eqn:asy}) 
This disagreement is evidently connected with the  
substantial difference between the gluon distributions obtained in 
different fits.
The gluon distribution was estimated in the analysis by 
Santiago and Yndurain as $\sim x^{-0.44}(1-x)^{8.1}$~\cite{Santiago:1999xb} 
with the momentum carried by gluons 
$0.752\pm0.014$ at $Q^2=12~{\rm GeV}^2$.
This is much larger than the typical value of $\sim 0.4$ obtained in 
the recent global fits of PDFs~\cite{Alekhin:2002fv,Martin:2002aw}.
Independent estimation of this quantity based on the results
of CDHS collaboration \cite{deGroot:1978hr} is $0.56\pm0.02$, also much 
lower than the result by Santiago and Yndurain. We comment that 
due to the value of $\alpha_{\rm s}$ is 
anti-correlated with the gluon distribution at small $x$, the 
increase in $\alpha_{\rm s}$ caused by 
the absence of the HT terms in that analysis is
compensated by its decrease due to enhanced gluons. 

\section{Conclusion} 

We also estimate theoretical errors connected with account of the heavy 
quarks contribution. These errors are more important than in our
previous determination of $\alpha_{\rm s}$ due to growing impact of 
the recent low-$x$ HERA data. The uncertainty 
in $\alpha_{\rm s}(M_{\rm Z})$ due to choice between 
the fixed-flavor-number and the variable-flavor-number 
factorization schemes is estimated as 0.0007.  
The variation of the $c$-quark mass by $0.25~{\rm GeV}$
leads to the variation in $\alpha_{\rm s}(M_{\rm Z})$ of 0.0006.
Other possible sources of theoretical errors,  
including uncertainties in the NNLO evolution kernel and 
in the strange quarks distribution, are not considered because they 
give much smaller effect. Summarizing all sources of theoretical errors we get
$$
\alpha^{\rm NNLO}_{\rm s}(M_{\rm Z})=0.1143
\pm 0.0014({\rm exp.})\pm 0.0013({\rm theor.}).
$$
This value is somewhat 
lower than the world average value of Ref.\cite{Bethke:2000ai}
but agrees with it within errors. 

{\bf Acknowledgments}

I am indebted to S.A.~Kulagin for careful reading of manuscript and 
valuable comments, D.V.~Shirkov, and A.~Vogt 
for stimulating discussions, and F.J.~Yndurain 
for communication of the details of his analysis.
I am grateful to the Theory Division of CERN, where the part of 
this work was done, for a hospitality.

\end{document}